\documentclass[%
reprint,
 amsmath,amssymb,
 pre,
 longbibliography,
]{revtex4-2}
\pdfoutput=1
\usepackage{siunitx}
\usepackage{esint}
\usepackage{subfigure}
\usepackage{graphicx}% Include figure files
\usepackage{dcolumn}% Align table columns on decimal point
\usepackage{bm}% bold math
\renewcommand{\Im}{\operatorname{Im}}

\begin{document}

\title{Dimensionless solutions of the wave equation}% Force line breaks with \\

\author{J. Blas}
\email{juabla@tel.uva.es}
\author{J.~L. Gutiérrez}
\author{E.~J. Abril}

\affiliation{
Department of Signal Theory, Communications and Telematics, University of Valladolid, 47011 Spain}

% \altaffiliation[Also at ]{}%Lines break automatically or can be forced with \\

%%\affiliation{%
%%Department of Signal Theory, Communications and Telematics, University of Valladolid, 47011 Spain.


%%\author{Charlie}
%\affiliation{}%
%\affiliation{}%
%%}%


\date{\today}% It is always \today, today,

%# 315 tribute worlf: focusing and defocusing aperture

\begin{abstract} 
  Plane waves are regarded as the general solution of the wave
  equation. However the plane wave expansion of standing waves by
  means of complex phasors leads to a theory in which the time
  coordinate does not receive the same treatment as the three space
  coordinates. An equal treatment is possible using our alternative
  approach built upon the dimensionless version of the wave
  equation. As a result, the usual standing wave solution written as
  sum of plane waves is just one of the available geometrical
  projections and therefore removes a part of the available
  information. The existence of these alternative projections and the
  constraints that they introduce, produce verifiable consequences. We
  present an experimental verification of one of this consequences by
  means of acoustic waves. In particular a resonant cavity is radiated
  from an external source through a squared aperture. The
  predicted flows of phase based on Pólya potentials allow us to find
  the direction of arrival without using temporal
  coordinates. Although this work is limited to the wave equation, the
  background concept is the relationship between space and time and
  therefore could have far reaching consequences in other physical
  models.
   
  % \begin{description}
% Secondary publications and information retrieval purposes.
% \item[PACS numbers]
% May be entered using the \verb+\pacs{#1}+ command.
% You may use the \texttt{description} environment to structure your abstract;
% use the optional argument of the \verb+\item+ command to give the category of each item. 
% \end{description}
\end{abstract}

\pacs{Valid PACS appear here}% PACS, the Physics and Astronomy
\maketitle

% \tableofcontents
\section{Introduction}
It is usual to think of the basic quaternionic imaginary units $i$,
$j$, $k$ \cite{Altmann.1986} as referring to three mutually
perpendicular (right-handed) axes in ordinary Euclidean
three-dimensional space \cite[p.~8]{Kravchenko.2003}. If we take the
real axis to represent the time coordinate, these quaternions would
describe a four-dimensional space-time. But it turns out that
quaternions are not appropiate for the description of spacetime in
this way because their natural quadratic form has an incorrect
signature for relativity theory \cite{RogerPenrose.2004}. So the
standard treatment is to identify a vector from $\mathbb{R}^3$ with a
purely vectorial quaternion, without real part. A complex quaternion
$q$ is an object of the form $q=q_0+q_1 i+q_2 j +q_3 k$, where
$q0$,$q1$,$q2$ and $q3$ are complex numbers. In this case, we must
establish the conmutation rule for the usual complex imaginary unit
with the quaternionic imaginary units. When both types of imaginary
units anticommute one obtains octonions
\cite{Kantor.1989,Dixon.1994,Ward.1997} which are nonassociative but
form a division algebra. In contrast, if they do not commute we obtain
complex quaternions. They enjoy the property of associativity but
there exist non-zero elements which do not have inverses
\cite[p.~11]{Kravchenko.2003}. Using complex quaternions and the
differential operator $D$ \cite{Moisil.1931,Gurlebeck.1997}:
\begin{equation}
D=i\frac{\partial}{\partial x} +j \frac{\partial}{\partial y} +k \frac{\partial}{\partial z},
\end{equation}
so $D^2=-\Delta$ would be the usual Laplace operator from
$\mathbb{R}^3$. It is possible to generalize one-dimensional complex
analysis in $\mathbb{R}^3$ and $\mathbb{R}^4$ by means of
hyperholomorphic functions \cite{Kravchenko.1996}. Modeling spatial
dimensions using imaginary units provides powerful insights that make
possible to treat rotational and divergence operators as different
aspects of the same operation, paving the way to the compact
expressions of vectorial algebra.

In contrast with the aforementioned approaches, in this work the
orthogonality between $i,j,k$ is used to model the orthogonality
between diferent components of the phase space instead of modeling
spatial dimensions. In order to establish the need for this
multidimensional phase space, consider the perfect mathematical
balance between space and time coordinates in the wave equation:
\begin{eqnarray}
  \lambda^2\frac{\partial^2p}{\partial x^2}=T^2\frac{\partial^2p}{\partial t^2}.
  \label{eq:waveeq}
\end{eqnarray}
Here $\lambda$ is the spatial period, $T$ is the temporal period, $x$
is one of the three independent variables of the Cartesian coordinate
system, $t$ is the time variable and $p(x,t)$ is a one-dimensional
wave. Both $\lambda$ and $T$ represent the completion of a cycle, and
can make space and time dimensionless quantities
\cite{HansPetterPetterLangtangen746}. In other words, their
relationship does not depend on the arbitrary choice of
units. Therefore space $\bar{x}=x/\lambda$ and time $\bar{t}=t/T$, as
dimensionless quantities, are clearly interchangeable in (\ref{eq:waveeq}):
\begin{eqnarray}
  \frac{\partial^2 p}{\partial \bar{x}^2}=\frac{\partial^2 p}{\partial \bar{t}^2}.
  \label{eq:waveeqdl}
\end{eqnarray}

The point that we will try to show in this work is that this
interchangeability is a fundamental property that cannot be violated
in any expression deduced from (\ref{eq:waveeqdl}). This
principle, hereinafter referred to as Space-Time Interchangeability
Principle (STIP), involves implicit constrains which are far from
being trivial.

If the time dependence is assumed to be in the form
$e^{-i2\pi \bar{t}}$ and is suppressed below, (\ref{eq:waveeqdl}) reduces to:
\begin{eqnarray}
  \frac{\partial^2 \tilde{p}}{\partial \bar{x}^2}+(2\pi)^2\tilde{p}=0,
\label{eq:wavesin}
\end{eqnarray}

where $\tilde{p}$ is a phasor. (\ref{eq:wavesin}) is a
second-order homogeneous linear ordinary differential equation with
constant coefficients and therefore its characteristic equation is a
quadratic with two roots \cite{Raza2018}, namely $\pm i2\pi$. The
general solution of (\ref{eq:wavesin}) is the sum of
progressive and regressive waves with complex coefficients $A$ and
$B$:
\begin{eqnarray}
  Ae^{-i 2\pi \bar{t}} e^{i 2\pi \bar{x}}+Be^{-i 2\pi \bar{t}} e^{-i 2\pi \bar{x}}.
  \label{eq:generalsolution}
\end{eqnarray}

The sign of $-i2\pi \bar{t}$ is not an issue when the time-reversal
counterpart of (\ref{eq:generalsolution}) is taken into
consideration. Equally harmless are in appearance an isolated
progressive wave ($Ae^{-i2\pi\bar{t}} e^{i2\pi\bar{x}}$) or an isolated
regressive ($Be^{-i2\pi \bar{t}} e^{-i 2\pi \bar{x}}$) wave. For any of
them, space and time produce rotations of their phasors in the complex
plane and therefore the STIP is apparently satisfied. However, a
contradiction arises if $A=B$ in (\ref{eq:generalsolution})
and therefore:
\begin{eqnarray}
  Ae^{-i2\pi\bar{t}} (e^{i2\pi\bar{x}}+ e^{-i2\pi \bar{x}})=  Ae^{-i2\pi\bar{t}}2\cos(2\pi \bar{x}),
\label{eq:standingwave}  
\end{eqnarray}
now the STIP is clearly violated, since $\bar{t}$ variations are
associated with a phasor rotation in the complex plane, while
$\bar{x}$ variations are not \cite[p.~11-14]{TristanNeedham1998}. There
are two options: either STIP is not always true or the solution with
$A=B$ has missing terms, since a fundamental feature has mysteriously
disappeared. The aim of this paper is to establish that the second
option is the correct one.

\section{Hypercomplex solutions of the wave equation}

The dimensionless version of the wave equation presented in
(\ref{eq:waveeqdl}) can be solved preserving the equal treatment of
space and time by means of hypercomplex phasors. A quaternion has
three imaginary units, namely $i,j,k$ while the scalar $1$ represents
the unit of the real part. These units obey the product rules given by
Hamilton: $i^2=j^2=k^2=-1$, $i=jk=-kj$, $j=ki=-ik$, $k=ij=-ji$. A
general superposition of progressive and regressive standing waves
would be written as:
\begin{equation}
  e^{j2\pi\bar{x}} A e^{-i2\pi\bar{t}} + e^{-j2\pi\bar{x}} B e^{-i2\pi\bar{t}},
  \label{eq:swquaternions}
\end{equation}
where $A, B \in \mathbb{H}$ are constant quaternions and the first
addend represents a standing wave whose source is the plane
$x=-\infty$ (progressive sense) while the second addend corresponds to
a standing wave whose source plane is $x=\infty$ (regressive
sense). Evident sources of standing waves are resonant cavities, for
example. This description guaranties an equal treatment of space and
time because both, $\bar{x}$ and $\bar{t}$ variations, always produce
rotations of the four-dimensional hypercomplex phasor. However, in
contrast with the plane waves in (\ref{eq:generalsolution}),
these rotations can happen in orthogonal hyperplanes instead of
sharing the same plane.
\begin{table}
\begin{tabular}{ c  c  c }
          & Space & Time \\
  \hline
  $\pm 1$ &  Antinode  & Antinode  \\
  $\pm i$ &  Antinode &  Node  \\
  $\pm j$ &  Node &  Antinode \\
  $\pm k$ &  Node & Node \\
\end{tabular} 
\caption{Symmetry for the different types of zero amplitude (nodes) of
  a standing wave}
\label{tab:zeros}
\end{table}

A fundamental consequence of (\ref{eq:swquaternions}) is that
each type of imaginary unit represents a different type of zero in the
amplitude of the wave. A closer look to a progressive standing wave
reveals that
\begin{eqnarray}
  e^{j2\pi\bar{x}} e^{-i2\pi\bar{t}}
&=&
  \left[\cos 2\pi\bar{x} + j \sin 2\pi\bar{x} \right]
  \left[\cos2\pi\bar{t} - i \sin 2\pi\bar{t} \right]=\nonumber \\
&=&
    \cos 2\pi\bar{x} \cos 2\pi\bar{t}
    -i \cos 2\pi\bar{x} \sin 2\pi\bar{t}  \nonumber \\
&+&  j \sin 2\pi\bar{x} \cos 2\pi\bar{t} +
    k \sin 2\pi\bar{x} \sin 2\pi\bar{t}.
    \label{eq:progsw}
\end{eqnarray}
The imaginary units $j$ and $k$ are simply necessary to represent the
perfect orthogonality between space an time as is sumarized in Table
\ref{tab:zeros}. Given that $e^{j 2\pi\bar{x}}e^{-i2\pi\bar{t}}$ is a
four-dimensional (4-D) sphere, due to its unit modulus, then any
change in the time coordinate or in the space coordinate represents a
rotation of the 4D-phasor. On the contrary, the usual phasor
$2Ae^{-i2\pi\bar{t}}\cos(2\pi\bar{x})$ becomes zero at a standing wave
node and lacks of amplitude, phase or energy. This is extremely
suspicious, because energy usually tends to fill the available
space. In contrast, by using 4D-phasors, the amplitude of the 4-D
phasor is constant in space-time using Eq. \ref{eq:progsw} and the
phase has no discontinuities.

A wave with the same appearance as a plane wave can be obtained adding
standing waves that are orthogonal between them in their space and
time rotations:
\begin{eqnarray}
  e^{j 2\pi \bar{x}} e^{-i 2\pi \bar{t}}+ e^{j 2\pi \bar{x}} ij e^{-i 2\pi \bar{t}}&=&  \nonumber \\
      \left[\cos 2\pi \bar{x} + j \sin 2\pi \bar{x} \right] && \left[\cos 2\pi \bar{t} - i \sin 2\pi \bar{t} \right]\nonumber \\
  + \left[\cos 2\pi \bar{x} + j \sin 2\pi \bar{x} \right] && \left[k\cos 2\pi \bar{t} -j\sin 2\pi \bar{t} \right]=\nonumber \\
=   \cos 2\pi \bar{x} \cos 2\pi \bar{t}&-& i \cos 2\pi \bar{x} \sin 2\pi \bar{t}  \nonumber \\
+ j \sin 2\pi \bar{x} \cos 2\pi \bar{t}&+& k \sin 2\pi \bar{x} \sin 2\pi \bar{t}  \nonumber \\
+   \sin 2\pi \bar{x} \sin 2\pi \bar{t}&+&  i \sin 2\pi \bar{x} \cos 2\pi \bar{t}  \nonumber \\
-  j \cos 2\pi \bar{x} \sin 2\pi \bar{t}&+& k \cos 2\pi \bar{x} \cos 2\pi \bar{t}  \nonumber \\
=  \cos( 2\pi \bar{x} - 2\pi \bar{t})  &+& i \sin(2\pi \bar{x} - 2\pi \bar{t}) \nonumber\\
 + j \sin(2\pi \bar{x} - 2\pi \bar{t}) &+& k \cos(2\pi \bar{x} - 2\pi \bar{t} ),     
\label{eq:apparentplainwaveasquaternion}
\end{eqnarray}

(\ref{eq:apparentplainwaveasquaternion}) is not just a plane
wave, because a plane wave would be only the term
$\cos(2\pi \bar{x} - 2\pi \bar{t}) + i \sin( 2\pi \bar{x}- 2\pi
\bar{t})$. This equation shows that the coefficients of $j$ and $k$
are different from zero. Therefore there are compensated forces underlying which
are not present in an ordinary plane wave expression. In contrast,
using the ordinary complex phasors for the same aim:
\begin{eqnarray}
\cos(2\pi\bar{x})e^{-i2\pi\bar{t}}+i \sin(2\pi\bar{x})e^{-i2\pi\bar{t}}=e^{i2\pi\bar{x}}e^{-i2\pi\bar{t}}.
\end{eqnarray}
A pure plane wave appears, without a description of the compensated
forces. Thus it provides again an incomplete representation: remember
that a compensated force is not the same as a nonexistent force.

In conclusion, using quaternionic standing waves every location in
space and every moment in time have their own phase, even standing
wave nodes. Every displacement in space or time produces always and
everywhere a rotation of the 4-D phasor without exceptions. From this
perspective, standing waves are at least as powerful in terms of
information as plane waves are. The usual standing wave phasor in the
complex plane is just a 2-D projection of this 4-D phasor and therefore
drops a lot of information.

Table \ref{tab:zeros} shows that $i$ and $j$ and $k$ are necessary to
obtain rotations that preserve an equal treatment of time and space
variations. However, the interpretation of $k$ is somewhat more
elusive. To cast some light on this issue, it must be said that the
action of substituting time for space and space for time can be
expressed as another rotation of $\pi/2$. Remember that $j=ki$ and
$i=jk$. For example, Nature employs this rotation to codify the
relationship between electric and magnetic fields so they also
preserve an equal treatment of time a space.

Maxwell equations relating electric and magnetic fields in absence of
sources are:
\begin{eqnarray}
\nabla\times{\mathbf{E}}&=&-\frac{\partial \mathbf{B}}{\partial t}  \\
\nabla\times{\mathbf{H}}&=& \frac{\partial \mathbf{D}}{\partial t}.  
\end{eqnarray}
Therefore a standing wave could be written also when time and space
are dimensionless as:
\begin{eqnarray}
j\mathbf{r}\times{\tilde{\mathbf{E}}}&=&i \tilde{\mathbf{B}} \\
j\mathbf{r}\times{\tilde{\mathbf{H}}}&=&-i \tilde{\mathbf{D}} 
\end{eqnarray}
where $\tilde{\mathbf{E}}$ and $\tilde{\mathbf{H}}$ are 4-D phasors. In consequence: 
\begin{eqnarray}
  \mathbf{r}\times{\tilde{\mathbf{E}}}&=&k \tilde{\mathbf{B}} \\
  \mathbf{r}\times{\tilde{\mathbf{H}}}&=&-k \tilde{\mathbf{D}} 
\end{eqnarray}
where $\mathbf{r}$ is the unit vector in the direction of propagation
and therefore $\mathbf{r}\times$ represents another $\pi/2$ rotation
in our three-dimensional space, while $k$ is a $\pi/2$ rotation in the
hypercomplex phase space. The presence of $\mathbf{r}\times$ is
clearly an implicit $\pi/2$ rotation in our three-dimensional space.

In particular, this rotations in our three-dimensional space fit very
well in the mathematics of geometric algebra. In the case of plane
waves it is customary\cite[p.~64-68]{ArthurJW.2011} to define a bivector
$I\mathbf{k}$, where $I$ is the unit pseudoscalar and $\mathbf{k}$ is
the propagation vector, from which we can generate a pseudoscalar
factor $I \mathbf{k}\cdot\mathbf{r}$ that will determine the phase of
some wave-front traveling along $\mathbf{k}$. As a bivector,
$I\mathbf{k}$ is actually associated with the plane of the wavefront,
whereas $\mathbf{k}$ points along the axis of propagation and is
therefore perpendicular to the wavefront. Solving the scalar wave
equation for the electromagnetic field $\boldsymbol{F}$, that is to
say, for $\mathbf{E}$ and $\mathbf{B}$ jointly yields:
\begin{equation}
\label{eq:emfield}
\boldsymbol{F} =(\mathbf{E}_0 + I \mathbf{B}_0) e^{I\left(\mathbf{k}\mathbf{r}-\omega t\right)}.
\end{equation}
The plane wave represented by (\ref{eq:emfield}) is inherently
circularly polarized. Taking $\mathbf{E}_0$ as lying in the
$I\mathbf{k}$ plane, then at any fixed point $\mathbf{r}$, the vectors
$\mathbf{E}$ and $\mathbf{B}$ rotate in quadrature about the
$\mathbf{k}$ axis with frequency $\omega$. The geometric algebra
language represents this spinning as the plane wave progresses as due
to a duality transformation rather than the usual kind of spatial
rotation\cite{DavidHestenes.2015}. However, here we deal with
quadratures of phase in 4-D which is a different concept. The combination of
both types of rotations --4-D rotation and spatial rotation--
under a unified mathematical language is out of the scope of this work.

If the momentum eigenstate for a plain wave is:
\begin{equation}
  e^{-i2\pi E t/h}e^{i 2\pi\mathbf{P}\cdot\mathbf{r}/h}
\end{equation}
where $\mathbf{P}$ is the spatial 3-momentum, $E$ is the energy, $h$
is the Planck’s constant, then for a standing wave the momentum
eigenstate should be:
\begin{equation}
  e^{-i2 \pi E t/h}e^{j 2\pi\mathbf{P}\cdot\mathbf{r}/h}.
\end{equation}
This would be another immediate consequence due to the wave-particle
duality which also preserves an equal treatment of time and space. A
detailed discussion is also out of the scope of this work.

\section{Complex potential for the wave flow based on phase}
Let us consider the wave equation in time-harmonic regime and three-dimensional space
(a generalization of (\ref{eq:wavesin})):
\begin{eqnarray}
   \label{eq:time-harmonic}
\nabla^2 U(\mathbf{r})=- (2\pi)^2 U(\mathbf{r}),
\end{eqnarray}
where $U$ represents a solution which is a superposition of
hypercomplex phasors at location $\mathbf{r}$. Time dependence
$e^{-i\omega t}=e^{-i2\pi \bar{t}}$ is implicit as usual. The projection
of $U(\mathbf{r})$ onto the $1i$ plane, also known as complex plane
would be:
\begin{eqnarray}
   \label{eq:general_sol_t}
U_t(\mathbf{r})=A_t(\mathbf{r})e^{i\psi_t(\mathbf{r})}  
\end{eqnarray}
where $A_t(\mathbf{r})$ is the resultant phasor amplitude and
$\psi_t(\mathbf{r})$ is the resultant phase, both of them on the
complex plane.  Alternatively, there is another proyection onto the $1j$
plane of the same solution $U$:
\begin{eqnarray}
   \label{eq:general_sol_s}
U_s(\mathbf{r})=A_s(\mathbf{r})e^{j\psi_s(\mathbf{r})}  
\end{eqnarray}
where $A_s(\mathbf{r})$ is the resultant phasor amplitude and
$\psi_s(\mathbf{r})$ is the resultant spatial phase, both of them on
the $1j$ plane. Both projections, $U_t$ and $U_s$, must satisfy the
wave equation separately. In the case of $U_t$ there is no doubt. In
contrast, $U_s$ must also satisfy the wave equation only if our
assumption is correct and the STIP holds:
\begin{eqnarray}
   \label{eq:time-harmonicUs}
\nabla^2 U_s(\mathbf{r})=- (2\pi)^2 U_s(\mathbf{r}),
\end{eqnarray}

The left-hand side
of (\ref{eq:time-harmonicUs}) can be expanded using the Laplacian
operator definition:
\begin{eqnarray}
   \label{eq:laplacianUs}
\nabla\cdot(\nabla U_s(\mathbf{r}))=-(2\pi)^2 U_s(\mathbf{r}). 
\end{eqnarray}
By replacing (\ref{eq:general_sol_s}) in (\ref{eq:laplacianUs}) and
omitting dependence on $\mathbf{r}$:
\begin{eqnarray}
   \label{eq:aplynabla}
\nabla\cdot(e^{j\psi_s}\nabla A_s+jA_se^{j\psi_s}\nabla\psi_s)=-(2\pi)^2 A_se^{j\psi_s}. 
\end{eqnarray}
Extracting common factor $A_se^{j\psi_s}$:
\begin{eqnarray}
   \label{eq:nablaapplyed}
\nabla\cdot\left[\left(\frac{\nabla A_s}{A_s}+j\nabla\psi_s\right) A_se^{j\psi_s}\right]=-(2\pi)^2 A_s e^{j\psi_s}. 
\end{eqnarray}
The term $\nabla A_s$ appears divided by $A_s$ and represents an
equivalent angular gradient of phase, $\nabla\phi_s$, which can be
added with $j\nabla\psi_s$. Analogously, in polar coordinates
$rd\theta=dr \implies d\theta=dr/r$. In other words, there exist a
more general phase that includes $\phi_s+j\psi_s=\ln A_s+j\psi_s$ in
its definition. The complex logarithm is uniquely defined (up to
constants) as the conformal mapping sending concentric circles with
constant $\phi_s$ to parallel lines. In other words, the logarithm is
an analytic function. The logarithmic mapping could be consulted for
reference in \cite[p.~100]{TristanNeedham1998}.

Our hypothesis is that $\phi_s+j\psi_s$ must provide also a conformal
mapping in our three-dimensional space in order to satisfy the
STIP. An analytic function of an analytic function is also analytic
\cite[p.~97]{polya1974}. Lets see in which way (\ref{eq:nablaapplyed}) is consistent with this hypothesis. The left
hand side of (\ref{eq:nablaapplyed}) after applying $\nabla$
and extracting common factor $U_s=A_se^{j\psi_s}$ is:
\begin{eqnarray}
    \label{eq:nablanablaapplyed}
\!\! \nabla^2U_s&=&  \left[\frac{A_s\nabla^2 A_s-\nabla A_s\cdot \nabla A_s}{A_s^2}+ j\nabla^2 \psi_s + \right. \nonumber\\
              &+&\!\!\left.\left(\frac{\nabla A_s}{A_s}+j\nabla\psi_s\right)\!\cdot\! \left(\frac{\nabla A_s}{A_s}+j\nabla\psi_s\right)\right] U_s
\end{eqnarray}
Due to the right-hand side of (\ref{eq:time-harmonicUs}),
$\nabla^2$ is an operator that can only change the amplitude of
$U(\mathbf{r})$ and not its phase, so the term in square brackets of
(\ref{eq:nablanablaapplyed}) has no imaginary part.  The
Cauchy-Riemann condition requires that both, the real and imaginary
parts of a differentiable complex function, such as $\phi_s+j\psi_s$
must satisfy Laplace's equation \cite[p.~95-96]{polya1974}:
\begin{eqnarray}
\Delta \psi_s= \Delta \phi_s=0.
\end{eqnarray}
As a consequence, the condition to cancel the imaginary part of (\ref{eq:nablanablaapplyed}) becomes:
\begin{eqnarray}
  \Im\left\{  \left(\frac{\nabla A_s}{A_s}+j\nabla\psi_s\right)\cdot\left(\frac{\nabla A_s}{A_s}+j\nabla\psi_s\right)\right\}=0
\end{eqnarray}
and therefore 
\begin{eqnarray}
  2j\frac{\nabla A_s}{A_s}\cdot\nabla\psi_s=0,
  \label{eq:ortogradient}
\end{eqnarray}
that is, $\nabla A_s$ and $\nabla \psi_s$ are orthogonal when they are
defined and are non-zero.

\begin{figure}
  \includegraphics{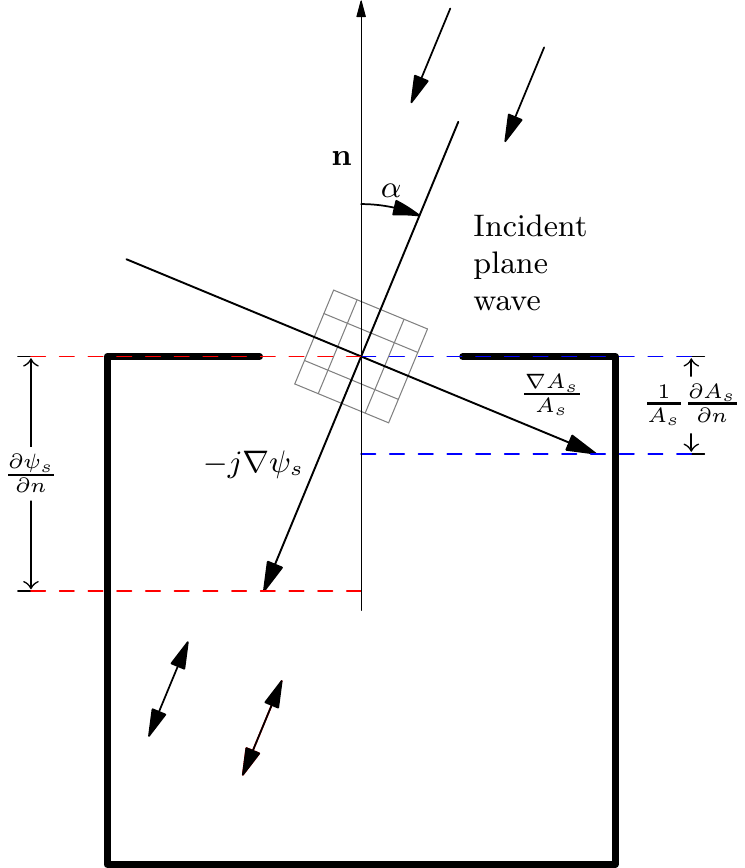}%
  \caption{Resonant cavity with aperture. An external plane wave
    crosses the aperture and impinges on a corner reflector which
    sends back a reflected wave, providing a maximum of spatial phase
    variation in this direction.  In contrast, the gradient of the
    phasor amplitude, which is another form of phase variation, has
    its maximum in the orthogonal direction.}\label{fig:propa}
\end{figure}

For example, consider a plane wave impinging on the aperture of a
cavity, as it is shown in Figure \ref{fig:propa}. The external plane
wave crosses the aperture and impinges on a corner reflector which
sends back a reflected wave, providing a maximum of $\psi_s$ variation
in this direction.  In contrast, the gradient of the phasor amplitude,
which is another form of phase variation, has its maximum in the
orthogonal direction. The gradients would follow straight lines in the
space and differential squares would become aligned with the
gradients. This conformal mapping has an associated Pólya complex
potential (using $j$ instead of $i$ because we are on the $1j$
plane). In this case we have a uniform flow of phase and the complex
potential is $f=2\pi(\bar{x}+j\bar{y})+\text{constant}$. Thus the
velocity of the fluid $f^\prime$ is everywhere constant and equals
$2\pi$ in terms of adimensional space coordinates. Using this
terminology, stream lines appear when
$\psi_s=0,\pm \kappa,\pm 2\kappa,\ldots$ while equipotencial lines
appear when $\phi_s=0,\pm \kappa,\pm 2\kappa$ where $\kappa$ is a
constant. The speed of the flow is represented by the crowding
together of the streamlines. No fluid can cross them. While
equipotential lines would represent a velocity potential.

The condition for the real part of $\nabla\cdot\nabla$ in (\ref{eq:nablanablaapplyed}) is also important:
\begin{eqnarray}
   \label{eq:nabla2comparison}
\nabla^2 U_s=\left(\frac{\nabla^2 A_s}{A_s}-(\nabla\psi_s)^2\right) U_s=-(2\pi)^2 U_s
\end{eqnarray}

If we have a Pólya complex potential, then $A_s$ must also
be harmonic, and therefore $\nabla^2 A_s=0$. This restriction together
with ($\ref{eq:nabla2comparison})$ implies that:
\begin{eqnarray}
  \label{eq:gpsi_s}
(\nabla\psi_s)^2=(2\pi)^2
\end{eqnarray}
In our example, a monochromatic plane wave is assumed to impinge on
the external surface of this aperture. In turn, the wave that crosses
the aperture bounces in the inner corner reflector inside the
cavity. As a consequence, there exists a strong reflection in the
opposite direction so the gradient of spatial phase $\nabla\psi_s$
reaches its maximum which equals the wave number of the medium,
satisfying (\ref{eq:gpsi_s}) and therefore forcing
$\nabla^2 A_s=0$ which is consistent with a Pólya complex potential.

\begin{figure*}
    \includegraphics{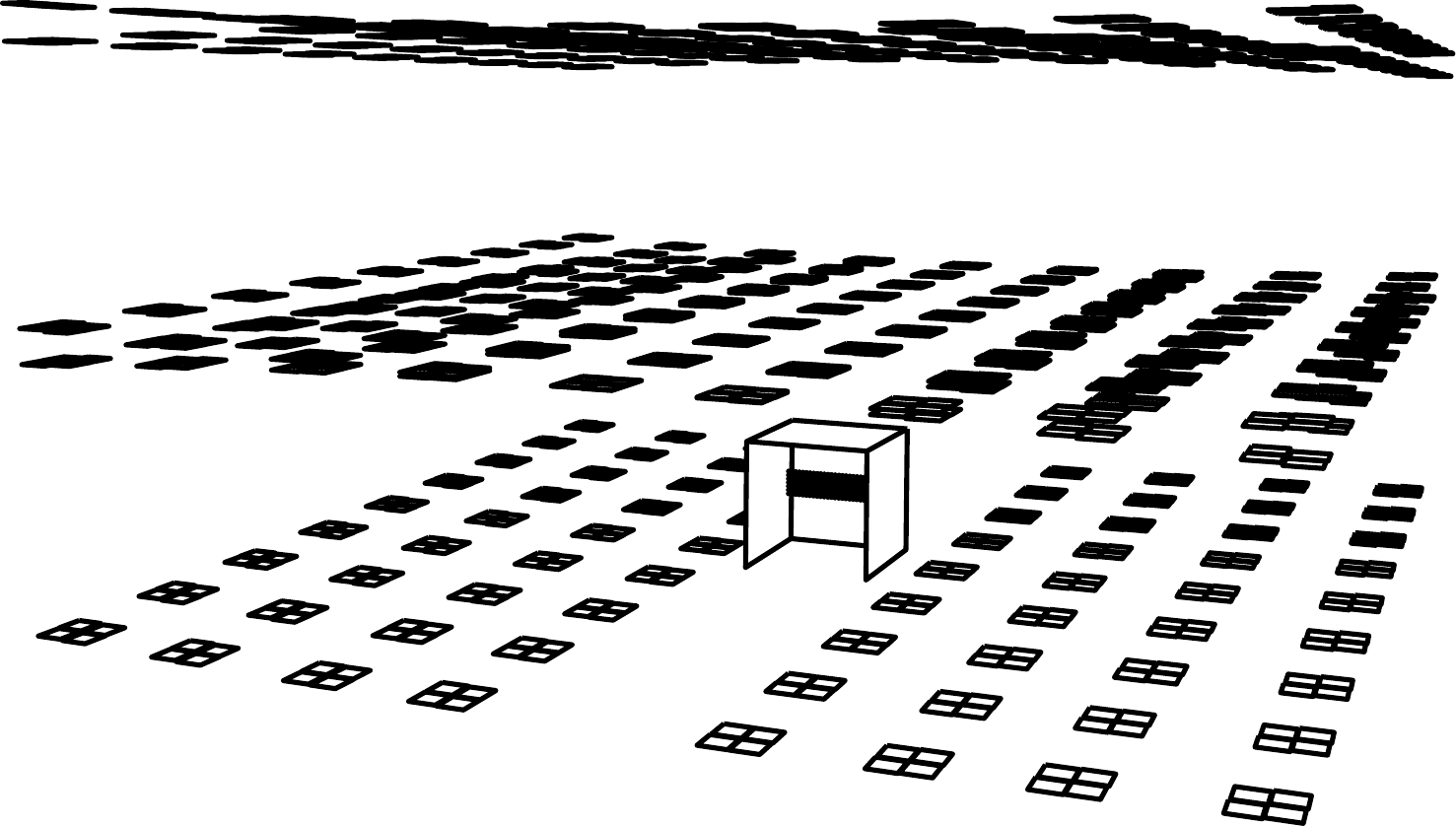}%
    \caption{Images of subapertures seen from the aperture center. The
      aperture wall and the cavity roof are not depicted to show the
      sampling points. Almost zenithal view. There are three vertical layers
      with 72, $81\times2$ and $81\times2$ aperture images, 396 images
      in total and 1584 subapertures.}\label{fig:images}
\end{figure*}

\section{Verifiable consequences using divergence theorem}

The Kirchhoff boundary conditions for the aperture have been found to
yield remarkably accurate results and are widely used in practice in
spite of their internal inconsistencies. So we consider unperturbed
plane waves on the aperture surface. Additionally, the aperture
provides an environment in which $\nabla A_s/A_s$ is expected to be
non-zero due to the presence of evanescent modes. As stated in
(\ref{eq:ortogradient}), $\nabla A_s$ and $\nabla\psi_s$ should
be orthogonal when they are non-zero.

Applying the divergence theorem to (\ref{eq:nablaapplyed}):
\begin{equation}
\oiint_S U_s \left(\frac{\nabla A_s}{A_s}+j\nabla\psi_s\right)\cdot d\mathbf{s}=
  -(2\pi)^2 \iiint_V U_s dv, \label{eq:flux_vect}
\end{equation}
where $S$ is a closed surface with differential surface normal
$d\mathbf{s}$, with modulus $ds$ and unit vector $\mathbf{n}$,
bounding the volume $V$. (\ref{eq:flux_vect}) in terms of
normal derivatives is therefore:
\begin{equation}
   \label{eq:flux_scalar}
  \oiint_S \!\! U_s \left(\frac{1}{A_s}\frac{\partial A_s}{\partial n}+j\frac{\partial\psi_s}{\partial n}\right) ds=
  -(2\pi)^2 \iiint_V U_s dv,
\end{equation}

According to the Helmholtz equivalency theorem\cite{Copson39}, an
aperture can be treated as a collection of secondary sources. In this
context, the Kirchhoff boundary conditions have been found to yield
remarkably accurate results and are widely used in practice in spite
of their internal inconsistencies. The Kirchhoff
solution is the arithmetic average of the two Rayleigh-Sommerfeld
solutions  which have
consistent boundary conditions \cite{Goodman68}. We consider plane
waves on the aperture surface $S^\prime$, neglecting fringing effects,
as first approximation. Under these conditions (\ref{eq:flux_scalar}) can be reduced to:
\begin{equation}
  \label{eq:flux_scalar_lhs}
  \left(\frac{1}{A_s} \frac{\partial A_s}{\partial n}+j\frac{\partial\psi_s}{\partial n}\right)
 \!\! \iint_{S^\prime} U_s ds^\prime = -(2\pi)^2\! \iiint_V\!\! U_s dv,
\end{equation}
where the first factor would be constant on the aperture surface
$S^\prime$ due to the unperturbed plane wave approximation, therefore
it can be expressed as follows:
\begin{eqnarray}
 \label{eq:FracIntUs}
 \left(\frac{1}{A_s} \frac{\partial
  A_s}{\partial n}+j\frac{\partial\psi_s}{\partial n}\right)=
  \frac{
  -(2\pi)^2 \iiint_V  U_s dv}
         {
  \iint_{S^\prime} U_s ds^\prime}.
\end{eqnarray}
The argument of the right hand side of (\ref{eq:FracIntUs}) can be simplified as follows:
\begin{eqnarray}
\label{eq:argument}
\arg\left\{
  \frac{
   \iiint_V U_sdv}
         {
  \iint_{S^\prime} U_s ds^\prime} \right\}
  &=&
 \arg\left\{  \iiint_V U_s dv \right\} \nonumber \\
      &-&
 \arg\left\{  \iint_{S^\prime} U_s  ds^\prime \right\},
\end{eqnarray}
where the limits of the last integral correspond to the aperture
surface $S^\prime$. On this surface, the unperturbed plane wave
approximation provides a linear phase distribution which is symmetric
respect to the center of the aperture $(0,0,0)$ and therefore:
\begin{equation}
\label{eq:centralpoint}
  \arg\left\{  \iint_{S^\prime} U_s  ds^\prime \right\}=\arg\left\{ U_s(0,0,0) \right\}=\psi_s(0,0,0). 
\end{equation}

The subtrahend of the right hand side of (\ref{eq:argument})
takes out the influence of the actual value of $\psi_s(0,0,0)$.
Therefore, without loss of generality, it will be considered hereafter
that: $\psi_s(0,0,0)=0$ at the aperture center.  The fact that
$\psi_s(0,0,0)=0$ is of great importance because for each point within
the cavity there exists only one phase $\psi_s(x,y,z)$ which is
consistent with this phase at the aperture center. Moreover
$\psi_s(x,y,z)$ can be calculated in advance using geometric
information. In particular, after using $\psi_s(0,0,0)=0$ at the
aperture center, the points inside the cavity have negative phase in
order to represent an outgoing flow of standing waves, which is
generated inside the cavity. Thus, with this new phase reference,
(\ref{eq:FracIntUs}) would be:
\begin{eqnarray}
 \label{eq:FracIntUs_minus}
 \left(\frac{1}{A_s} \frac{\partial
  A_s}{\partial n}-j\frac{\partial\psi_s}{\partial n}\right)=
  \frac{
  -(2\pi)^2 \iiint_V A_se^{-j\psi_s} dv}
         {
  \iint_{S^\prime} A_se^{-j\psi_s} ds^\prime}.
\end{eqnarray}

The projection of $\frac{\nabla A_s}{A_s}-j\nabla\psi_s$ onto the unit
normal $\mathbf{n}$ is shown in Figure \ref{fig:propa} and is given
by:
\begin{eqnarray}
\frac{1}{A_s}\frac{\partial A_s}{\partial n}-j\frac{\partial\psi_s}{\partial n}=2\pi(-\sin\alpha -j\cos\alpha), \label{eq:sincos}
\end{eqnarray}
where the sense of $\nabla\psi_s$ is consistent with increasing phases
in the sense of the outgoing flow and the sense of $\nabla{A_s}$ is
due to an exponential decay of the amplitude in the outward
direction. The real and the imaginary parts of this number represent
the two legs of the same right triangle with hypotenuse $2\pi$, since
we are assuming a uniform flow through the aperture. Therefore we
obtain a verifiable consequence: this right triangle should determine
the angle of incidence $\alpha$ of the original plane wave which comes
from the external source:
\begin{equation}
  \alpha=\frac{\pi}{2}-\arg\left\{
      \frac{\iiint_V A_se^{-j\psi_s} dv}
      {\iint_{S^\prime} A_se^{-j\psi_s} ds^\prime}
      \right\}. \label{eq:alpha}
\end{equation}

\begin{figure}
  \includegraphics{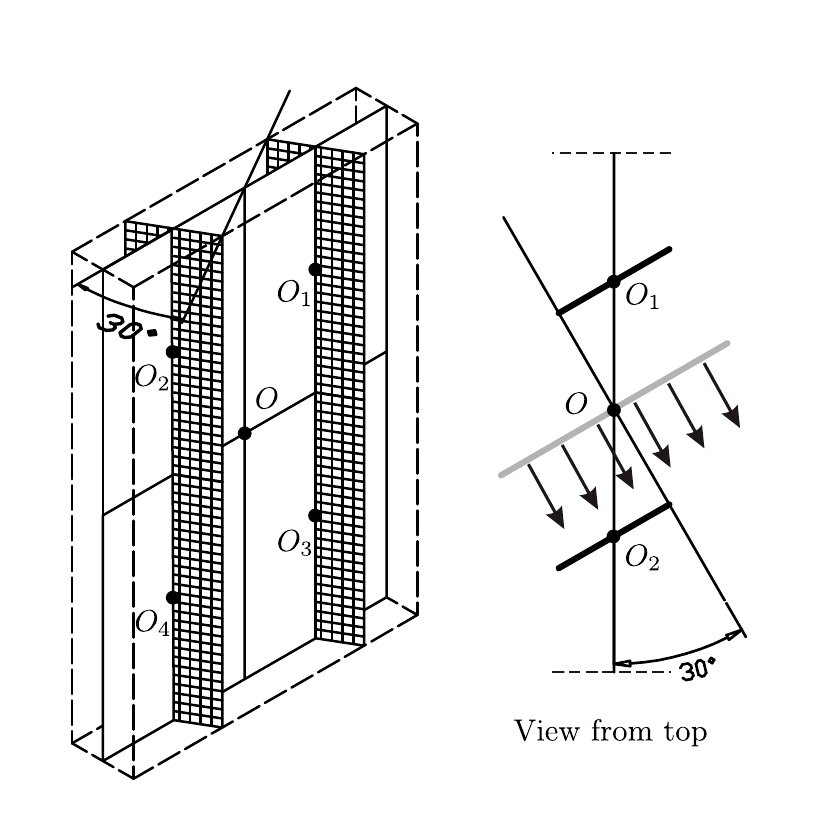}%
  \caption{Contributions from an aperture can be divided
    into several subwindows.}\label{fig:subwindows}
\end{figure}

As stated before, measurements inside the cavity must be performed in
far field conditions. This requirement can be addressed treating the
aperture as a collection of subapertures
\cite{juabla2008,juabla2009}. For example, Figure \ref{fig:subwindows}
shows an aperture which has been divided in four subapertures. The
zenithal view under far field approximation illustrates parallel
propagation vectors from secondary sources radiating towards the
target point. The superposition of the secondary sources at the
subapertures is equivalent to the superposition of the secondary
sources on the original aperture. In this example, the waves departing
from secondary sources at subapertures $1$ and $3$ have longer path
lengths than if they were to depart from the original
aperture. However, in turn, subapertures $2$ and $4$ have a shorter
pathlength. So the error of phase of each subaperture is compensated
globally because under unperturbed plane wave approximation all the
secondary sources have the same amplitude.  As a conclusion, under far
field conditions, the original aperture with null phase at its
aperture center $O$ is equivalent to the four subapertures with null
phase at their subaperture centers $O_1$, $O_2$, $O_3$ and $O_4$.

The presence of walls in the cavity has an important influence on the
amplitude distribution inside the cavity. Walls behavior can be
modeled by acoustic images of the real aperture that take into account
reflections in the walls. In summary we substitute the effect of the
walls by the effect of these equivalent apertures. Our aim is to find
$\psi_s$ for each sampling point inside the cavity provided that
$\psi_s=0$ at the aperture center. In order to find this phase, we
need to identify which images must be taken into account. This
approach simplifies the numerical solution because wall influence is
treated using essentially the same tools that are needed to solve an
isolated aperture.

\begin{figure}
\centering
  \includegraphics{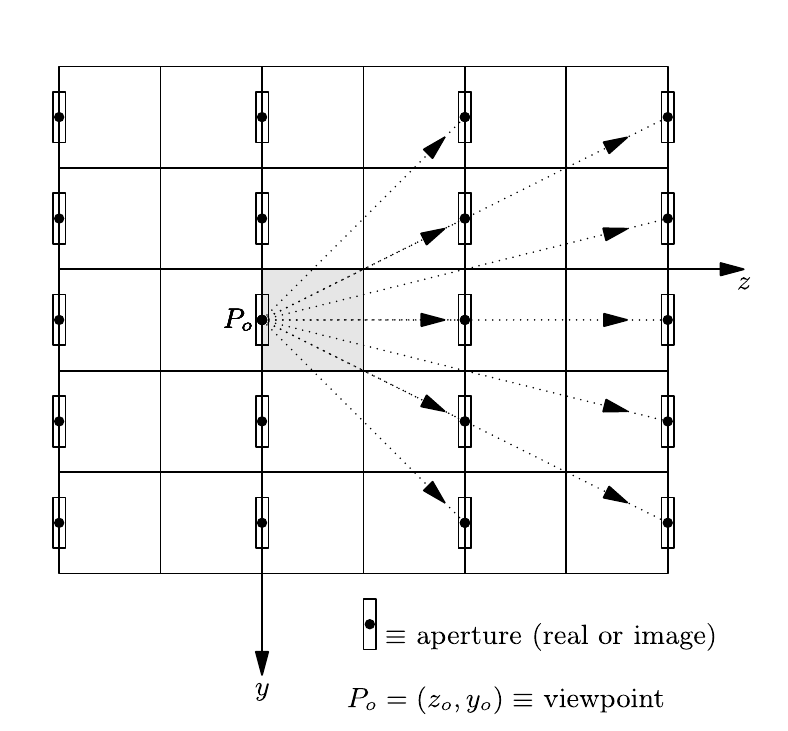}%
  \caption{Only some acoustic images are source of waves that leave the
    cavity. Images with $z>0$ can send energy through the aperture
    because they are seen from the real aperture center $P_o$ looking
    inward.}\label{fig:images-example}
\end{figure}

However, not all the acoustic images of the real aperture in the six
walls can be considered as sources of standing waves that impinge on
the aperture surface in the outgoing direction. In the example shown
in figure \ref{fig:images-example}, only those images with $z>0$, or
equivalently those apertures whose center is visible from the real
aperture center viewpoint while looking inward should be taken into
account. Those images with $z<0$ represent energy reflected in the
aperture wall which does not leave the cavity and does not contribute
to the outgoing flux of spatial waves. Only energy coming from the
rest of the walls is eligible for modeling outgoing standing waves.
\begin{figure}
\centering
    \includegraphics{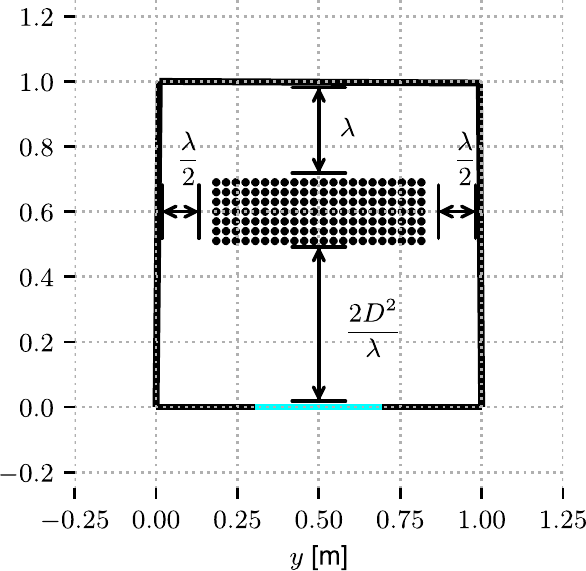}%
  \caption{Sampling grid (height above the cavity ground of \SI{0.545}{\meter}).}\label{fig:sampling}
\end{figure}

\section{Experimental verification}
The experimental verification of (\ref{eq:time-harmonicUs})
and (\ref{eq:alpha}), which are backed up by the STIP, has been carried
out by means of a resonant cavity with a squared aperture centered on
the frontal wall of the cavity. A high-fidelity tweeter located
outside the cavity radiated a pure tone with
$\lambda=\SI{0.3}{\meter}$ through this aperture. The aperture was
surrounded by an acoustic adsorbing material to prevent waves from
entering the cavity other than through the aperture. The cavity was
approximately a cube of edge \SI{1}{\meter} although three of the four
vertical faces were slightly rotated so that opposite faces were not
parallel and consequently the interference pattern inside the cavity
were more chaotic. A microphone (Earthworks M23) with a typical
sensitivity of \SI{34}{\milli\volt\per\pascal} and uniform polar
pattern was used. Its unique circuitry excludes the transconductance
of the input FET from the overall gain structure. This means the
sensitivity remains very stable when the microphone is subjected to
variations in ambient temperature. The microphone was connected to a
phantom power supply (Triton Audio True Phantom) with low noise
components to achieve low distortion and improve the signal-to-noise
ratio. The microphone signal was made available to a dynamic signal
analyzer (HP 35670A) by means of an impedance transformer with
\SI{110}{\ohm} input and \SI{75}{\ohm} output. For each sample, the
analyzer averaged 5 time records using power averaging mode.

The microphone was carried on a rail guided vehicle to take spatial
samples of acoustic intensity level inside the cavity. This vehicle
was driven by three stepper motors, controlled by a laptop which dealt
with the synchronization of measurements and displacements. The
external acoustic source was also moved by means of another
rail-guided vehicle.  The source begins its motion in front of the
center of the aperture, at a distance of \SI{1.2}{\meter}. Its
trajectory was parallel to the aperture surface using discrete
increments of \SI{5}{\centi\meter}. A servomotor was used to point the
tweeter to the center of the aperture after each step
automatically. Temperature and humidity conditions were also monitored
in order to detect variations during the experiment. In addition a
second microphone was placed outside the cavity at a fixed location to
verify the absence of other external acoustic sources.

\begin{figure*}
  \includegraphics{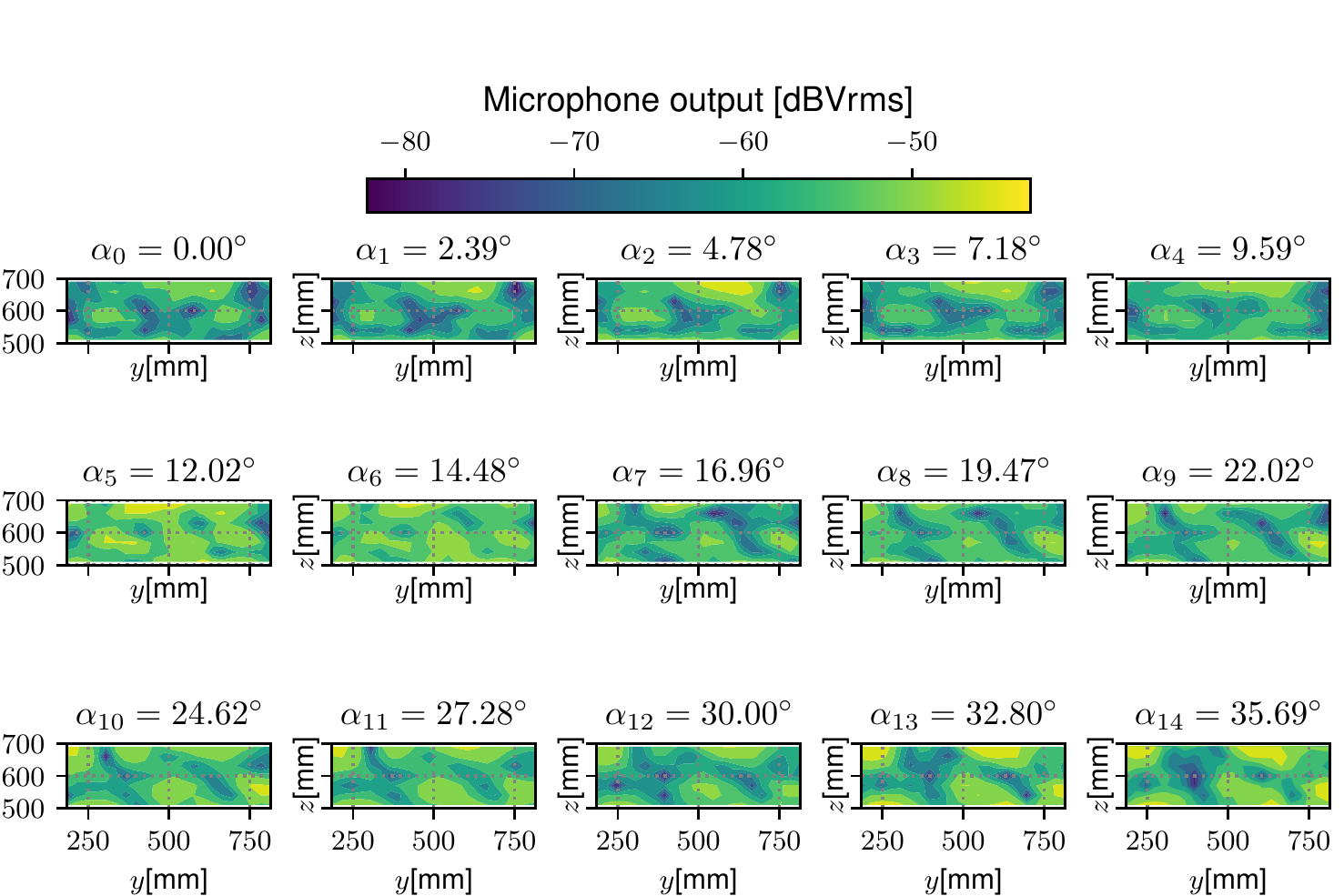}%
  \caption{Measurements inside the cavity. Source location
    $y_s=0.5-0.05n\,\text{[m]}, z_s=1.2\,\text{[m]}$ with
    $n=0,1,2,\ldots,14$. Angle of incidence
    $\alpha_n$.}\label{fig:measurements}
\end{figure*}

Under ordinary conditions, there are temperature and humidity
variations in the cavity over time that modify the wavelength of the
acoustic field during the experiment. In order to avoid those
wavelength variations, the temporal frequency $f$ of the wave
generator was tuned for each sample in order to keep the wavelength
$\lambda$ constant. This is possible using $f=v/\lambda$, where $f$ is
the temporal frequency, $v$ is the speed of sound and $\lambda$ is a
constant. In particular, we calculated the speed of sound in humid air
as a function of the instantaneous temperature, relative humidity and
pressure \cite{Cramer93}. The saturation vapor pressure was taken from
\cite{Davis92}.
%, and we considered a mole fraction of carbon dioxide of 0.0004.

The sampling grid to measure sound intensity inside the cavity had 154
measurement points, as shown in Figure \ref{fig:sampling}. We treat
the aperture as a collection of 4 sub-apertures. Each sub-aperture is
a square with side \SI{19.5}{\centi\meter}, each one satisfying the
conventional far-field criteria. Therefore the distance from the
sampling grid to the sub-apertures is:
\begin{equation}
  \frac{2D^2}{\lambda}\approx \SI{0.51}{\meter},
\end{equation}
where the diagonal of each sub-aperture is
$D\approx\SI{0.276}{\meter}$ and $\lambda=\SI{0.3}{\meter}$ is the
wavelength.
%This approximation is consistent with the measurements, as
%shown in Figure \ref{fig:measurements}, where the hatched areas
%represent measurement points which are too close to the aperture to
%fulfill far field conditions and therefore exhibit a particular
%spatial field distribution.

The microphone and the rail-guided vehicle inside the cavity occupy
some space, so there are points near the walls which are out of
reach. Anyway, the measurements taken too close to the cavity walls
are unreliable, as a result, some points must be left out of the
sampling grid. The distance between samples is
$0.1\lambda=\SI{3}{\centi\meter}$, which is the usual distance
employed to retain enough information about the spatial distribution
of the fields.

In general, in order to implement the divergence theorem, we would
need a three-dimensional sampling volume inside the cavity instead of
a two-dimensional sampling area. However, in our experimental setup,
the direction of incidence was restricted to a horizontal plane, and
under those conditions, the spatial distribution of the fields in a
horizontal plane near the center of the cavity is assumed to contain
enough information to estimate the angle of incidence. This assumption
is based on our previous experiments \cite{juabla2008,juabla2009}. In
those previous works we found that two-dimensional sampling areas in
far field conditions provide an electric field envelope Cumulative
Distribution Function (CDF) which identifies statistically the angle
of incidence, independently of the precise location of each sample and
regardless of not covering all the space near the cavity walls.

The lateral walls of the experimental cavity are not parallel as shown
in Figure \ref{fig:sampling}. Thus lateral images are seen from the
real aperture center in contrast with the previous example in Figure
\ref{fig:images-example}. In turn, those lateral images have very
close images due to the wall of the real aperture. But given that this
second row of images are behind the real aperture, they are not seen
directly from the real aperture.  However, their reflections on the
back wall, opposite to the aperture wall are necessary to represent
certain geometrical conditions of the phase of the outgoing signal due
to the lack of parallelism between walls. This is the reason why some images have a
close duplicate in the layers of images which represent the back wall
effect in Figure \ref{fig:images}, which includes all the images taken
into account. It is not necessary to include all the images which are
seen from the real aperture. Although there is an infinite number of
them, not all the images are equally important. For example those that
are far away have a weaker influence and their geometrical description
tends to be more prone to cumulative errors.

Our experimental approximation to (\ref{eq:alpha}) for the
angle of incidence $\alpha$ is:
\begin{equation}
\label{eq:experimental}
\alpha=\frac{\pi}{2} - \arg\left\{ \sum_{n \mathop{=} 1}^{N}  \vert U_s(n) \vert\, e^{j\psi_n}\right\}
\end{equation}
where $N=154$ is the total number of sampling points,
$\vert U_s(n) \vert$ is the amplitude measured in the $n^{\rm{th}}$
sampling point and $\psi_n$ is the phase calculated assuming
$\psi_s=0$ at the supaperture center and the same amplitude in every
image. In particular, the expression for $\psi_n$ can be calculated in
anticipation using only geometrical data and is calculated only once:
\begin{equation}
  \label{eq:refphase}
  \psi_n=\arg\left\{\sum_{m \mathop{=} 1}^{M} \frac{e^{-j\frac{2\pi}{\lambda} d_{nm}}}{d_{nm}}\right\},
\end{equation}
where $\lambda=\SI{0.3}{\meter}$ is the wavelength, $d_{mn}$ is the
distance between the $n^{\rm{th}}$ sampling point and the center of
the $m^{\rm{th}}$ subaperture image. $M=1584$ is the total number of
images which are taken into account (shown in Figure \ref{fig:images}).

\begin{table*}
  \centering
  \begin{tabular}{|r|ccccccccccccccc|}
\hline 
Ideal result  &  0.0 &  2.4 &  4.8 &  7.1 & 9.5 & 11.8 & 14.0 & 16.3 & 18.4 & 20.6 & 22.6 & 24.6 & 26.6 & 28.4 & 30.3 \\
\hline
Regression 1 &  0.5 &  2.8 &  5.1 &  7.3 &  9.6 & 11.8 & 14.0 & 16.2 & 18.3 & 20.3 & 22.3 & 24.3 & 26.2 & 28.0 & 29.7 \\
Regression 2 & 2.1 &  4.4 &  6.6 &  8.8 & 11.0 & 13.1 & 15.2 & 17.3 & 19.3 & 21.3 & 23.2 & 25.1 & 26.9 & 28.7 & 30.4 \\
Raw result 1 & 4.0 & 7.7 & 7.5 & 6.3 &  2.0 &  2.0 &  6.5 & 15.3 & 25.5 & 29.0 & 26.9 & 23.7 & 23.5 & 27.5 & 29.0 \\
Raw result 2 & 4.5 & 14.5 & 12.3 & 6.6 &  4.0 & 3.5 & 4.8 & 13.1 & 21.8 & 26.0 & 25.3 & 25.5 & 28.5 & 32.2 & 31.0 \\
    \hline
  \end{tabular} \caption{Numerical comparison between the angles of incidence in degrees, the raw angles obtained from (\ref{eq:experimental}) and the estimated angles using a linear regression of the raw angles. }\label{tab:predictions}
\end{table*}

\section{Results and conclusions}

\begin{figure}
  \includegraphics{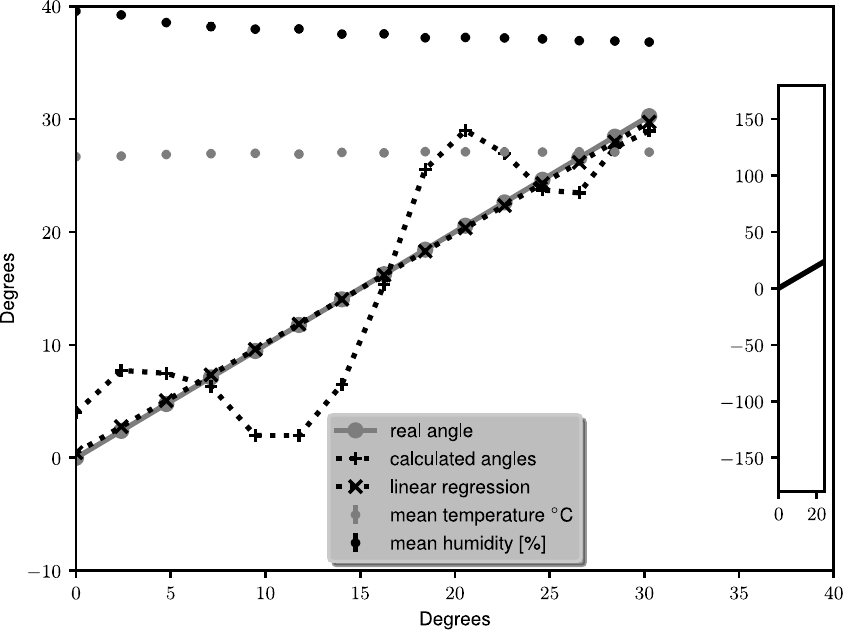}%
  \caption{Comparison between the real angle of incidence and the
    calculated ones. The prediction is finally fitted using a linear
    regression. The inset shows a general view covering all the
    possible values.}\label{fig:regression}
\end{figure}

The experimental results are summarized in Figure
\ref{fig:regression}. We provide a comparison between the real angles
of incidence and those predicted by means of (\ref{eq:experimental}). Our sampling region does not cover all the
space inside the cavity and there are several approximations,
therefore it is not possible to obtain an accurate prediction for each
isolated location of the external source. However, by using a linear
regression we can compensate the errors.

The fitted prediction follows with accuracy the external source
displacements of $\lambda/6$ using a $3.3\lambda\times 3.3\lambda$
cavity. All the calculated angles were the result of considering a
fixed set of 243936 hypercomplex numbers that provide a description of
the geometry of the cavity in terms of the location of the sampling
points.

If (\ref{eq:experimental}) were not related to the angle of
incidence we would expect a much more chaotic distribution of
predictions. In fact, the argument of a random complex number could
have any value in $[\ang{-180},\ang{180})$. In contrast, instead of a
chaotic distribution we obtain a set of points oscillating around a
regression line which predicts fairly close the ideal
dataset. Moreover, the same experiment was repeated, with different
profiles of temperature and humidity. Due to the environmental
variations, each realization of the same experiment provided a
different prediction, however they were quite similar as can be seen
in Table \ref{tab:predictions}.

In conclusion, we have provided a mathematical formalism and a
physical interpretation for solving the adimensional version of the
wave equation in compliance with the Space-Time Interchangeability
Principle, which in turn, is strongly backed by other well proven
physical theories such as Relativity. This formalism has been
validated by providing experimental data which confirm its
predictions. The solution is based on quaternions and predicts
additional forms of phase which are necessary to obtain a more
complete description of the wave phenomena. Evanescent waves which are
presently not fully understood play an important role in the
experimental model. The concept of phase flow in terms of Pólya
potential, helps to integrate the wave phenomena with other
related physical phenomena, such as electrostatics and fluid
motion. Although this work is limited to the wave equation, the
background concept is the relationship between space and time and
therefore could have far reaching consequences in other physical
models.

%\nocite{*}
\bibliography{ms}% Produces the bibliography via BibTeX.

\providecommand{\noopsort}[1]{}\providecommand{\singleletter}[1]{#1}%
\begin{thebibliography}{21}%
\makeatletter
\providecommand \@ifxundefined [1]{%
 \@ifx{#1\undefined}
}%
\providecommand \@ifnum [1]{%
 \ifnum #1\expandafter \@firstoftwo
 \else \expandafter \@secondoftwo
 \fi
}%
\providecommand \@ifx [1]{%
 \ifx #1\expandafter \@firstoftwo
 \else \expandafter \@secondoftwo
 \fi
}%
\providecommand \natexlab [1]{#1}%
\providecommand \enquote  [1]{``#1''}%
\providecommand \bibnamefont  [1]{#1}%
\providecommand \bibfnamefont [1]{#1}%
\providecommand \citenamefont [1]{#1}%
\providecommand \href@noop [0]{\@secondoftwo}%
\providecommand \href [0]{\begingroup \@sanitize@url \@href}%
\providecommand \@href[1]{\@@startlink{#1}\@@href}%
\providecommand \@@href[1]{\endgroup#1\@@endlink}%
\providecommand \@sanitize@url [0]{\catcode `\\12\catcode `\$12\catcode
  `\&12\catcode `\#12\catcode `\^12\catcode `\_12\catcode `\%12\relax}%
\providecommand \@@startlink[1]{}%
\providecommand \@@endlink[0]{}%
\providecommand \url  [0]{\begingroup\@sanitize@url \@url }%
\providecommand \@url [1]{\endgroup\@href {#1}{\urlprefix }}%
\providecommand \urlprefix  [0]{URL }%
\providecommand \Eprint [0]{\href }%
\providecommand \doibase [0]{https://doi.org/}%
\providecommand \selectlanguage [0]{\@gobble}%
\providecommand \bibinfo  [0]{\@secondoftwo}%
\providecommand \bibfield  [0]{\@secondoftwo}%
\providecommand \translation [1]{[#1]}%
\providecommand \BibitemOpen [0]{}%
\providecommand \bibitemStop [0]{}%
\providecommand \bibitemNoStop [0]{.\EOS\space}%
\providecommand \EOS [0]{\spacefactor3000\relax}%
\providecommand \BibitemShut  [1]{\csname bibitem#1\endcsname}%
\let\auto@bib@innerbib\@empty
%</preamble>
\bibitem [{\citenamefont {Altmann}(1986)}]{Altmann.1986}%
  \BibitemOpen
  \bibfield  {author} {\bibinfo {author} {\bibfnamefont {S.}~\bibnamefont
  {Altmann}},\ }\href@noop {} {\emph {\bibinfo {title} {Rotations, quaternions
  and double groups}}}\ (\bibinfo  {publisher} {Clarendon Press},\ \bibinfo
  {year} {1986})\ pp.\ \bibinfo {pages} {9--28}\BibitemShut {NoStop}%
\bibitem [{\citenamefont {Kravchenko}(2003)}]{Kravchenko.2003}%
  \BibitemOpen
  \bibfield  {author} {\bibinfo {author} {\bibfnamefont {V.~V.}\ \bibnamefont
  {Kravchenko}},\ }\href@noop {} {\emph {\bibinfo {title} {Applied Quaternionic
  Analysis}}}\ (\bibinfo  {publisher} {Heldermann},\ \bibinfo {year}
  {2003})\BibitemShut {NoStop}%
\bibitem [{\citenamefont {Penrose}(2004)}]{RogerPenrose.2004}%
  \BibitemOpen
  \bibfield  {author} {\bibinfo {author} {\bibfnamefont {R.}~\bibnamefont
  {Penrose}},\ }\href@noop {} {\emph {\bibinfo {title} {The Road to Reality: A
  Complete Guide to the Laws of the Universe}}}\ (\bibinfo  {publisher} {Random
  House},\ \bibinfo {year} {2004})\ pp.\ \bibinfo {pages}
  {201--203}\BibitemShut {NoStop}%
\bibitem [{\citenamefont {Kantor}\ and\ \citenamefont
  {Solodovnikov}(1989)}]{Kantor.1989}%
  \BibitemOpen
  \bibfield  {author} {\bibinfo {author} {\bibfnamefont {I.~L.}\ \bibnamefont
  {Kantor}}\ and\ \bibinfo {author} {\bibfnamefont {A.~S.}\ \bibnamefont
  {Solodovnikov}},\ }\href@noop {} {\emph {\bibinfo {title} {Hypercomplex
  numbers}}}\ (\bibinfo  {publisher} {Springer},\ \bibinfo {year} {1989})\ pp.\
  \bibinfo {pages} {41--51}\BibitemShut {NoStop}%
\bibitem [{\citenamefont {Dixon}(1994)}]{Dixon.1994}%
  \BibitemOpen
  \bibfield  {author} {\bibinfo {author} {\bibfnamefont {G.}~\bibnamefont
  {Dixon}},\ }\href@noop {} {\emph {\bibinfo {title} {Division algebras:
  octonions, quaternions, complex numbers, and the algebraic design of
  physics}}}\ (\bibinfo  {publisher} {Springer},\ \bibinfo {year} {1994})\ pp.\
  \bibinfo {pages} {31--49}\BibitemShut {NoStop}%
\bibitem [{\citenamefont {Ward}(1997)}]{Ward.1997}%
  \BibitemOpen
  \bibfield  {author} {\bibinfo {author} {\bibfnamefont {J.}~\bibnamefont
  {Ward}},\ }\href@noop {} {\emph {\bibinfo {title} {Quaternions and Cayley
  Numbers}}}\ (\bibinfo  {publisher} {Springer},\ \bibinfo {year} {1997})\ pp.\
  \bibinfo {pages} {105--213}\BibitemShut {NoStop}%
\bibitem [{\citenamefont {Moisil}\ and\ \citenamefont
  {Théodoresco}(1931)}]{Moisil.1931}%
  \BibitemOpen
  \bibfield  {author} {\bibinfo {author} {\bibfnamefont {G.}~\bibnamefont
  {Moisil}}\ and\ \bibinfo {author} {\bibfnamefont {N.}~\bibnamefont
  {Théodoresco}},\ }\bibfield  {title} {\bibinfo {title} {Functions
  holomorphes dans l'espace},\ }\href@noop {} {\bibfield  {journal} {\bibinfo
  {journal} {Mathematica Cluj}\ }\textbf {\bibinfo {volume} {5}},\ \bibinfo
  {pages} {142} (\bibinfo {year} {1931})}\BibitemShut {NoStop}%
\bibitem [{\citenamefont {Gürlebeck}\ and\ \citenamefont
  {Sprössig}(1997)}]{Gurlebeck.1997}%
  \BibitemOpen
  \bibfield  {author} {\bibinfo {author} {\bibfnamefont {K.}~\bibnamefont
  {Gürlebeck}}\ and\ \bibinfo {author} {\bibfnamefont {W.}~\bibnamefont
  {Sprössig}},\ }\href@noop {} {\emph {\bibinfo {title} {Quaternionic and
  Clifford Calculus for Physicists and Engineers}}}\ (\bibinfo  {publisher}
  {John Wiley \& Sons},\ \bibinfo {year} {1997})\ pp.\ \bibinfo {pages}
  {37--74}\BibitemShut {NoStop}%
\bibitem [{\citenamefont {Kravchenko}\ and\ \citenamefont
  {Shapiro}(1996)}]{Kravchenko.1996}%
  \BibitemOpen
  \bibfield  {author} {\bibinfo {author} {\bibfnamefont {V.~V.}\ \bibnamefont
  {Kravchenko}}\ and\ \bibinfo {author} {\bibfnamefont {M.}~\bibnamefont
  {Shapiro}},\ }\href@noop {} {\emph {\bibinfo {title} {Integral
  Representations for Spatial Models of Mathematical Physics}}}\ (\bibinfo
  {publisher} {Longman},\ \bibinfo {year} {1996})\ pp.\ \bibinfo {pages}
  {1--11}\BibitemShut {NoStop}%
\bibitem [{\citenamefont {Langtangen}\ and\ \citenamefont
  {Pedersen}(2016)}]{HansPetterPetterLangtangen746}%
  \BibitemOpen
  \bibfield  {author} {\bibinfo {author} {\bibfnamefont {H.~P.~P.}\
  \bibnamefont {Langtangen}}\ and\ \bibinfo {author} {\bibfnamefont {G.~K.}\
  \bibnamefont {Pedersen}},\ }\href@noop {} {\emph {\bibinfo {title} {Scaling
  of Differential Equations}}}\ (\bibinfo  {publisher} {Springer},\ \bibinfo
  {year} {2016})\ p.~\bibinfo {pages} {71}\BibitemShut {NoStop}%
\bibitem [{\citenamefont {Tahir-Kheli}(2018)}]{Raza2018}%
  \BibitemOpen
  \bibfield  {author} {\bibinfo {author} {\bibfnamefont {R.}~\bibnamefont
  {Tahir-Kheli}},\ }\href@noop {} {\emph {\bibinfo {title} {Ordinary
  Differential Equations: Mathematical Tools for Physicists}}},\ \bibinfo
  {edition} {1st}\ ed.\ (\bibinfo  {publisher} {Springer},\ \bibinfo {year}
  {2018})\BibitemShut {NoStop}%
\bibitem [{\citenamefont {Needham}(1998)}]{TristanNeedham1998}%
  \BibitemOpen
  \bibfield  {author} {\bibinfo {author} {\bibfnamefont {T.}~\bibnamefont
  {Needham}},\ }\href@noop {} {\emph {\bibinfo {title} {Visual Complex
  Analysis}}}\ (\bibinfo  {publisher} {Clarendon Press},\ \bibinfo {year}
  {1998})\BibitemShut {NoStop}%
\bibitem [{\citenamefont {Arthur}(2011)}]{ArthurJW.2011}%
  \BibitemOpen
  \bibfield  {author} {\bibinfo {author} {\bibfnamefont {J.~W.}\ \bibnamefont
  {Arthur}},\ }\href@noop {} {\emph {\bibinfo {title} {Understanding Geometric
  Algebra for Electromagnetic Theory}}}\ (\bibinfo  {publisher} {John Wiley \&
  Sons},\ \bibinfo {year} {2011})\BibitemShut {NoStop}%
\bibitem [{\citenamefont {Hestenes}\ and\ \citenamefont
  {Lasenby}(2015)}]{DavidHestenes.2015}%
  \BibitemOpen
  \bibfield  {author} {\bibinfo {author} {\bibfnamefont {D.}~\bibnamefont
  {Hestenes}}\ and\ \bibinfo {author} {\bibfnamefont {A.}~\bibnamefont
  {Lasenby}},\ }\href@noop {} {\emph {\bibinfo {title} {Space-Time Algebra}}}\
  (\bibinfo  {publisher} {Birkhäuser},\ \bibinfo {year} {2015})\ pp.\ \bibinfo
  {pages} {31--33}\BibitemShut {NoStop}%
\bibitem [{\citenamefont {Polya}\ and\ \citenamefont
  {Latta}(1974)}]{polya1974}%
  \BibitemOpen
  \bibfield  {author} {\bibinfo {author} {\bibfnamefont {G.}~\bibnamefont
  {Polya}}\ and\ \bibinfo {author} {\bibfnamefont {G.}~\bibnamefont {Latta}},\
  }\href@noop {} {\emph {\bibinfo {title} {Complex Variables}}}\ (\bibinfo
  {publisher} {John Wiley \& Sons},\ \bibinfo {year} {1974})\BibitemShut
  {NoStop}%
\bibitem [{\citenamefont {Baker}\ and\ \citenamefont
  {Copson}(1939)}]{Copson39}%
  \BibitemOpen
  \bibfield  {author} {\bibinfo {author} {\bibfnamefont {B.~B.}\ \bibnamefont
  {Baker}}\ and\ \bibinfo {author} {\bibfnamefont {E.~T.}\ \bibnamefont
  {Copson}},\ }\href@noop {} {\emph {\bibinfo {title} {The Mathematical Theory
  of Huygens' Principle}}}\ (\bibinfo  {publisher} {Oxford, University Press},\
  \bibinfo {year} {1939})\ p.~\bibinfo {pages} {23}\BibitemShut {NoStop}%
\bibitem [{\citenamefont {Goodman}(1968)}]{Goodman68}%
  \BibitemOpen
  \bibfield  {author} {\bibinfo {author} {\bibfnamefont {J.}~\bibnamefont
  {Goodman}},\ }\href@noop {} {\emph {\bibinfo {title} {Introduction to Fourier
  Optics Goodman}}}\ (\bibinfo  {publisher} {McGraw-Hill},\ \bibinfo {year}
  {1968})\ pp.\ \bibinfo {pages} {40--51}\BibitemShut {NoStop}%
\bibitem [{\citenamefont {Blas}\ \emph {et~al.}(2008)\citenamefont {Blas},
  \citenamefont {Fern\'andez}, \citenamefont {Lorenzo}, \citenamefont {Abril},
  \citenamefont {Mazuelas}, \citenamefont {Bahillo},\ and\ \citenamefont
  {Bullido}}]{juabla2008}%
  \BibitemOpen
  \bibfield  {author} {\bibinfo {author} {\bibfnamefont {J.}~\bibnamefont
  {Blas}}, \bibinfo {author} {\bibfnamefont {P.}~\bibnamefont {Fern\'andez}},
  \bibinfo {author} {\bibfnamefont {R.~M.}\ \bibnamefont {Lorenzo}}, \bibinfo
  {author} {\bibfnamefont {E.~J.}\ \bibnamefont {Abril}}, \bibinfo {author}
  {\bibfnamefont {S.}~\bibnamefont {Mazuelas}}, \bibinfo {author}
  {\bibfnamefont {A.}~\bibnamefont {Bahillo}},\ and\ \bibinfo {author}
  {\bibfnamefont {D.}~\bibnamefont {Bullido}},\ }\href@noop {} {\bibfield
  {journal} {\bibinfo  {journal} {Progress In Electromagnetics Research}\
  }\textbf {\bibinfo {volume} {85}},\ \bibinfo {pages} {147} (\bibinfo {year}
  {2008})}\BibitemShut {NoStop}%
\bibitem [{\citenamefont {Blas}\ \emph {et~al.}(2009)\citenamefont {Blas},
  \citenamefont {Lorenzo}, \citenamefont {Fern\'andez}, \citenamefont {Abril},
  \citenamefont {Bahillo}, \citenamefont {Mazuelas},\ and\ \citenamefont
  {Bullido}}]{juabla2009}%
  \BibitemOpen
  \bibfield  {author} {\bibinfo {author} {\bibfnamefont {J.}~\bibnamefont
  {Blas}}, \bibinfo {author} {\bibfnamefont {R.~M.}\ \bibnamefont {Lorenzo}},
  \bibinfo {author} {\bibfnamefont {P.}~\bibnamefont {Fern\'andez}}, \bibinfo
  {author} {\bibfnamefont {E.~J.}\ \bibnamefont {Abril}}, \bibinfo {author}
  {\bibfnamefont {A.}~\bibnamefont {Bahillo}}, \bibinfo {author} {\bibfnamefont
  {S.}~\bibnamefont {Mazuelas}},\ and\ \bibinfo {author} {\bibfnamefont
  {D.}~\bibnamefont {Bullido}},\ }\href@noop {} {\bibfield  {journal} {\bibinfo
   {journal} {Progress In Electromagnetics Research}\ }\textbf {\bibinfo
  {volume} {91}},\ \bibinfo {pages} {101} (\bibinfo {year} {2009})}\BibitemShut
  {NoStop}%
\bibitem [{\citenamefont {Cramer}(1993)}]{Cramer93}%
  \BibitemOpen
  \bibfield  {author} {\bibinfo {author} {\bibfnamefont {O.}~\bibnamefont
  {Cramer}},\ }\bibfield  {title} {\bibinfo {title} {The variation of the
  specific heat ratio and the speed of sound in air with temperature, pressure,
  humidity, and co2 concentration},\ }\href {https://doi.org/10.1121/1.405827}
  {\bibfield  {journal} {\bibinfo  {journal} {The Journal of the Acoustical
  Society of America}\ }\textbf {\bibinfo {volume} {93}},\ \bibinfo {pages}
  {2510} (\bibinfo {year} {1993})}\BibitemShut {NoStop}%
\bibitem [{\citenamefont {Davis}(1992)}]{Davis92}%
  \BibitemOpen
  \bibfield  {author} {\bibinfo {author} {\bibfnamefont {R.~S.}\ \bibnamefont
  {Davis}},\ }\bibfield  {title} {\bibinfo {title} {Equation for the
  determination of the density of moist air (1981/91)},\ }\href
  {https://doi.org/10.1088/0026-1394/29/1/008} {\bibfield  {journal} {\bibinfo
  {journal} {Metrologia}\ }\textbf {\bibinfo {volume} {29}},\ \bibinfo {pages}
  {67} (\bibinfo {year} {1992})}\BibitemShut {NoStop}%
\end{thebibliography}%
\end{document}